\documentclass[twocolumn,aps,prl,superscriptaddress,floatfix]{revtex4-2}
\usepackage{graphicx,subfigure}
\usepackage{amsmath,amssymb,bm}
\usepackage{mathtools}
\usepackage{multirow}
\usepackage{makecell}
\usepackage{textcomp}
\usepackage{float}
\usepackage{color}
\usepackage[normalem]{ulem}
\usepackage{dsfont}
\usepackage{mathrsfs}
\usepackage{braket}
\usepackage{transparent}
\usepackage{xcolor}
\usepackage{hhline}
\usepackage{graphicx}
\usepackage{pdfpages}
\usepackage{listings}
\usepackage{xcolor}
\usepackage[export]{adjustbox}
\usepackage[dvipsnames]{xcolor}
\usepackage{hyperref}
\usepackage{physics}
\usepackage{graphicx}
\usepackage{lineno} 

\newcommand{\CrI}{\ensuremath{\mathrm{CrI}_{3}}}
\newcommand{\moire}{moir\'{e}}

\newcommand{\YCY}[1]{\textcolor{black}{#1}}
\newcommand{\YY}[1]{\textcolor{black}{#1}}
\newcommand{\YCYNEW}[1]{\textcolor{black}{#1}}
\newcommand{\YYNEW}[1]{\textcolor{black}{#1}}

\newcommand{\Modified}[1]{ \textcolor{black}{#1}}

\makeatletter
\AtBeginDocument{\let\LS@rot\@undefined}
\makeatother

\graphicspath{ {figures/} }

\makeatletter
\let\saved@includegraphics\includegraphics
\AtBeginDocument{\let\includegraphics\saved@includegraphics}
\makeatother

\begin{document}

\title{Universal Reconstruction of Complex Magnetic Profiles with Minimum Prior Assumptions}

\author{
Changyu Yao$^{1,*}$, Yue Yu$^{1,*}$, Yinyao Shi$^{1,*}$, Ji-In Jung$^{2}$, Zoltán Váci$^{3}$, Yizhou~Wang$^{1}$, Zhongyuan~Liu$^{1}$, Chuanwei Zhang$^{1,4}$, Sonia Tikoo-Schantz$^{2}$, Chong Zu$^{1,4,5,^\dag}$
\\
\medskip
\normalsize{$^{1}$Department of Physics, Washington University, St. Louis, MO 63130, USA}\\
\normalsize{$^{2}$Department of Geophysics, Stanford University, Stanford, CA 94305, USA}\\
\normalsize{$^{3}$Department of Earth and Planetary Sciences, Washington University, St. Louis, MO 63130, USA}\\
\normalsize{$^{4}$Center for Quantum Leaps, Washington University, St. Louis, MO 63130, USA}\\
\normalsize{$^{5}$Institute of Materials Science and Engineering, Washington University, St. Louis, MO 63130, USA}\\
\normalsize{$^*$These authors contributed equally to this work}\\
\normalsize{$^\dag$To whom correspondence should be addressed; E-mail: zu@wustl.edu}\\
}

\begin{abstract}
Understanding intricate magnetic structures in materials is essential for advancing materials science, spintronics, and geology. 
Recent developments of quantum-enabled magnetometers, such as nitrogen-vacancy (NV) centers in diamond, have enabled direct imaging of magnetic field distributions across a wide range of magnetic profiles.
However, reconstructing the magnetization from an experimentally measured magnetic field map is a complex inverse problem, further complicated by measurement noise, finite spatial resolution, and variations in sample-to-sensor distance. 
In this work, we present a novel and efficient GPU-accelerated method for reconstructing spatially varying magnetization density from measured magnetic fields with minimal prior assumptions. 
We validate our method by simulating diverse magnetic structures under realistic experimental conditions, including multi-domain ferromagnetism and magnetic spin textures such as skyrmion, anti-skyrmion, and meron.
Experimentally, we reconstruct the magnetization of a micrometer-scale Apollo lunar mare basalt (sample 10003,184) and a nanometer-scale twisted double-trilayer \CrI.
The basalt exhibits soft ferromagnetic domains consistent with previous paleomagnetic studies, whereas the \CrI\ system reveals a well‑defined hexagonal magnetic \moire\ superlattice.
Our approach provides a versatile and universal tool for investigating complex magnetization profiles, paving the way for future quantum sensing experiments.
\end{abstract}

\date{\today}

\maketitle

\section{Introduction}

Characterizing complex magnetic profiles is essential for understanding the fundamental mechanisms of magnetic materials and geological samples, as well as for advancing next-generation spintronics technology~\cite{vzutic2004spintronics,barman2020magnetization,glenn2017micrometer}.
Quantum-enabled magnetometers have made significant advancements in probing the local magnetic field distribution with unprecedented accuracy and resolution~\cite{degen2017quantum,aslam2023quantum}.
Existing quantum systems include, for example, atomic vapors~\cite{li2018serf,patton2014all,romalis2011atomic,fabricant2023build,robbes2006highly,cai2020herriott,huang2016single}, nitrogen-vacancy (NV) centers in diamond~\cite{maletinsky2012robust, casola2018probing,hsieh2019imaging,aslam2023quantum,glenn2015single,barry2020sensitivity,gross2016direct, levine2019principles, bhattacharyya2024imaging, davis2023probing, barry2020sensitivity} and spin defects in two-dimensional materials~\cite{gottscholl2020initialization,gottscholl2021room, healey2023quantum, gong2023coherent, vaidya2023quantum, gong2024isotope}, and superconducting quantum interference devices (SQUIDs)~\cite{kleiner2004superconducting,cohen1972magnetoencephalography,enpuku1999detection,cohen1970magnetocardiograms,drung1990low}. 
Their diverse applications enables the exploration of magnetic properties in various materials~\cite{casola2018probing,xu2023recent,liu2022magnetic,fragkos2022magnetic,dovzhenko2018magnetostatic}, biological processes~\cite{aslam2023quantum,ichkitidze2015magnetic,zhang2021toward,le2013optical}, geological samples~\cite{fu2020high,glenn2017micrometer,lima2016ultra,lima2009obtaining,steele2023paleomagnetic}, and beyond.

Accurate reconstruction of magnetic profiles from the measured field map presents a challenging inverse problem.
In reality, the presence of experimental measurement noise, finite imaging resolution, and sample-to-sensor distance further complicate the reconstruction process.
Current methods include inverse Fourier transformation~\cite{roth1989using,broadway2020improved,dubois2022untrained}, moment reconstruction~\cite{fu2020high,ghasemifard2017current,lima2016ultra}, and machine learning techniques\Modified{~\cite{reed2024machine, Broadway2025}}. Each of these approaches has its own limitations. The inverse Fourier transformation is commonly employed for reconstructing current density rather than magnetization. 
When used to solve for magnetization, the approach requires the prior assumption of a fixed axis along which the magnetization is constrained, a requirement that is often inapplicable in various scenarios~\cite{fragkos2022magnetic,liu2022magnetic,dovzhenko2018magnetostatic,saha2022observation}. 
The moment reconstruction method limits the sample to a point magnetic multiple, losing the spatial structure of the underlying magnetization. 
Meanwhile, the machine learning approach~\cite{Broadway2025} requires extensive computational resources and large training datasets to accurately capture the full range of physical phenomena, \Modified{and the results may depend highly on initial guess, which limits the reconstruction universality for unknown configurations}.

\begin{figure*}[ht]
    \centering
    \includegraphics[width=0.99\textwidth]{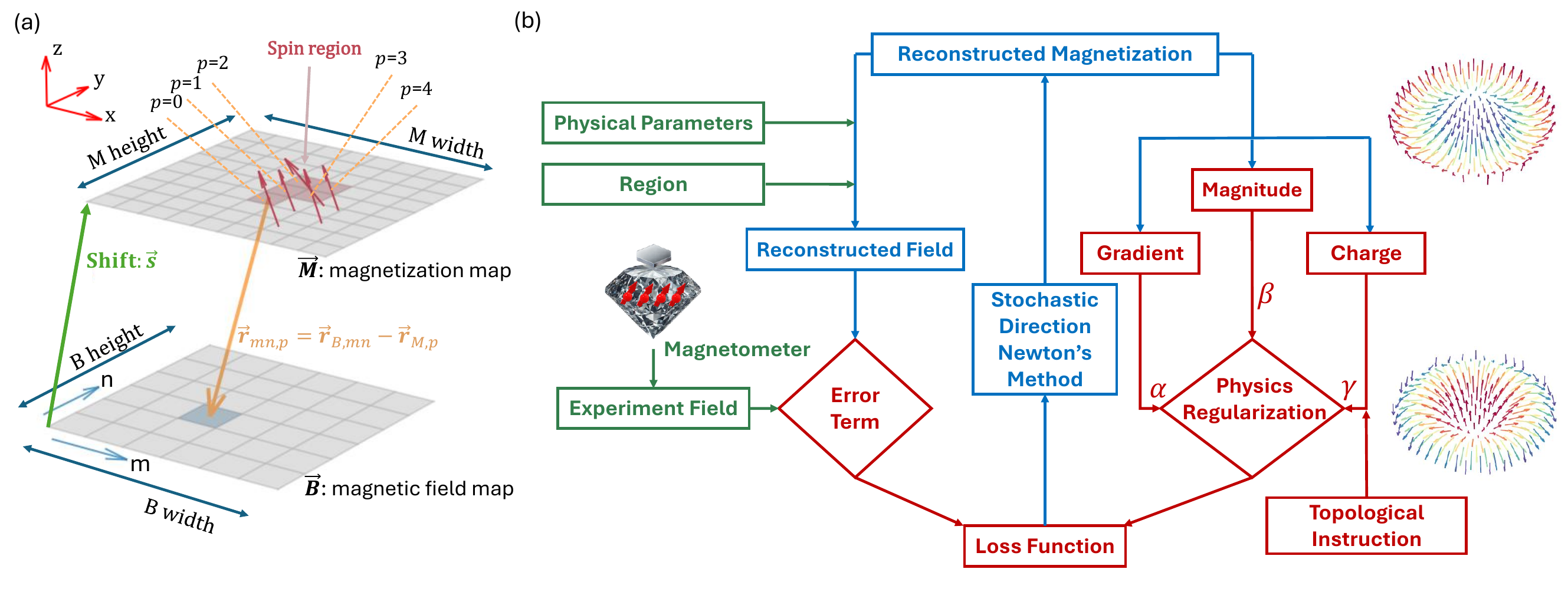}
    \flushleft
    \caption{(a) Schematics of calculating magnetic fields on grid $(m,n)$ \YCY{generated by spins in allocated region. We discretize the magnetization map into a 2D grid with additional rearrangement of the source region into an array of dipole moment vectors $\vec{\mathbf{m}}_p$ with a single index} $\mathbf{p} = (p_1, p_2, \ldots, p_N)$, and the magnetic field is obtained by summing the contributions from each element of $\mathbf{p}$. \YCY{The green vector denotes the relative position of the origin of the magnetization map shifted from magnetic field map. The two maps are assumed to be parallel.} (b) A demonstration of the complete reconstruction process is as follows: the green sections represent the experimental procedures, the red sections represent our customizable loss function, and the blue sections represent the computational procedures. Any quantum sensor capable of measuring 2D magnetic field maps would be suitable for the design. The illustration on the left side of the figure shows a quantum diamond magnetometer as an example. The spin textures displayed on the right side of the figure illustrate a Néel-type anti-skyrmion at the top and a skyrmion at the bottom.}
    \label{fig:diag}
\end{figure*}

\begin{figure*}[ht]
    \centering
    \includegraphics[width=0.99\textwidth]{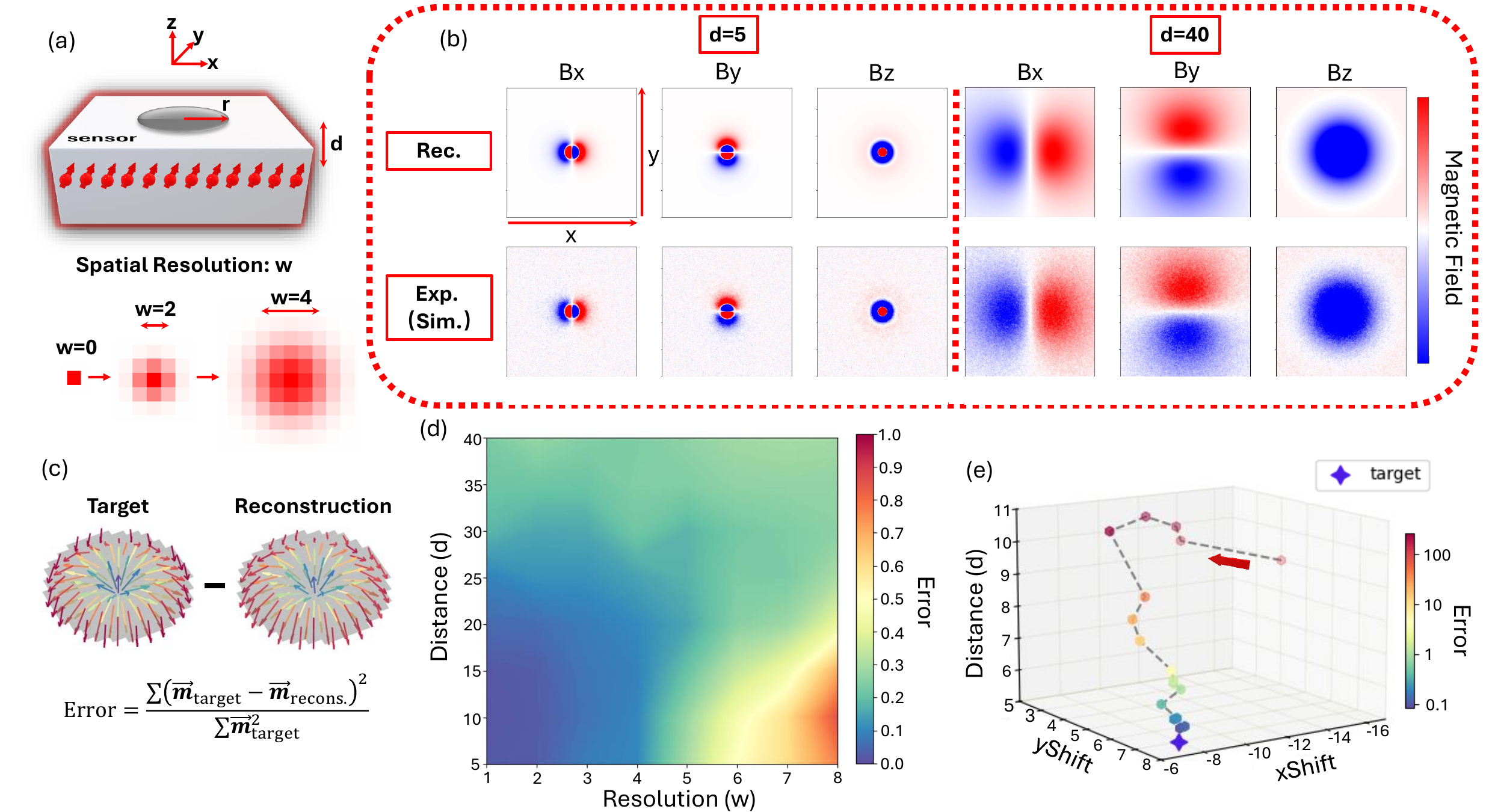}
    \flushleft
    \caption{Case study of Néel-type skyrmion. (a) \YY{Schematic of the computational experiment setup, where $d$ is the sample-to-sensor distance, $r$ the skyrmion radius, and $w$ the full width at half maximum (FWHM) of the 2D Gaussian convolution modeling the experimental resolution. All parameters are expressed in dimensionless values normalized to pixel length. }The skyrmion radius is fixed at $r = 8$ throughout this study. (b) \YY{Comparison of simulated and reconstructed magnetic field maps at small and large sample-to-sensor distances $d$. All maps use the same color scale and dimensions ($144 \times 144$ pixels).} (c) Demonstration of the error calculation. Here, the reconstructed magnetization is obtained with parameters $d = 5$, $w = 2$, and SNR $= 5$. (d) Color map showing reconstruction error across different combinations of spatial resolution ($w$) and sample-to-sensor distance ($d$). \YCY{(e) Visualization of parameters optimization with the shift vector $\vec{s}$, where each point represents a single optimization run, and the star denotes the target shift vector.}} 
\label{fig:skyrmion}
\end{figure*}

In this work, we present three main results.
First, we develop a GPU-accelerated computational model capable of reconstructing the full spatially varying magnetization from three-dimensional magnetic field maps (Figure \ref{fig:diag}).
To guide the optimization process for various complex magnetic profiles existing in nature, we introduce a customizable physics-informed loss function and implement the stochastic directional Newton method to efficiently drive the reconstruction process. 
Second, we validate our method through the simulations of various magnetic structures, including multi-domain ferromagnetism and topological spin textures such as skyrmion, anti-skyrmion, and meron under realistic experimental conditions (Figure \ref{fig:skyrmion} and Figure \ref{fig:case}).
In particular, by carefully examining the errors of reconstruction against finite measurement noise, spatial resolution, and sample-to-sensor distance, we find the optimal parameter regime for guiding future quantum sensing experiments.
\Modified{Experimentally}, we apply our model to analyze the measured magnetic field maps from a lunar rock (Apollo lunar mare basalt sample 10003,184) acquired using a wide-field quantum diamond microscope (Figure~\ref{fig:rock}).
\Modified{
Furthermore, we validate the model's ability to resolve nanoscale magnetic structures by reconstructing the magnetization of a twisted double trilayer \CrI\ with known \moire\ superlattice constant (Figure~\ref{fig:moire})\YYNEW{, using magnetic field map measured by a scanning NV microscope~\cite{song2021moire}}.
}

\section{Description of the Model}

The essence here is to solve a magnetic inverse problem. The magnetic field generated by a magnetic dipole is:
\begin{equation}
    \vec{\mathbf{B}} = \frac{\mu_0}{4\pi r^3}\left( \vec{\mathbf{m}} - 3\cdot \hat{r}(\hat{r}\cdot\vec{\mathbf{m}})\right)\,\text{,}
\end{equation}
where $\vec{r} = \vec{r}_{B} - \vec{r}_{m}$. 
\YCY{
We discretize both the field and magnetization maps onto 2D grids (Figure \ref{fig:diag}(a)). 
Indices $m$ and $n$ are used to mark x and y positions on the magnetic field grid.
We define the magnetization region, shown in red in Figure~\ref{fig:diag}(a), based on prior knowledge of the magnetic source (e.g., the white light image of the sample).
The magnetization is rearranged into a 1D array of dipole moment vectors with index, $p$, so that we can tackle any desired shape of sample using least amount of storage and computational resources.
Within this region, a magnetic dipole is placed at the center of each pixel, while pixels outside are assigned to zero moment and therefore do not contribute to the total field. 
At each step of the process, we adjust \YCYNEW{each component of the} moment \YCYNEW{independently} of each assigned dipole according to the workflow shown in Figure~\ref{fig:diag}(b).
}

Using superposition principle, the $i$th component of $\vec{\mathbf{B}}$ on grid $(m,n)$ can be calculated as 
\begin{equation}
    \begin{split}
        B_{mn}^{i} &= \frac{\mu_0}{4\pi}\sum_{p,j}\frac{1}{r_{mn,p}^3}\left(\delta_{ij}m_{p}^{j} - 3\hat{r}_{mn,p}^{i}\hat{r}_{mn,p}^{j}m_{p}^j\right) \\
        &= A_{mn,p}^{i,j} m_{p}^{j}\,\text{.}
    \end{split}
\end{equation}
$\mathbf{A}$ is the coefficient matrix:
\begin{equation}
    A_{mn,p}^{i,j} = \frac{\mu_0}{4\pi}\times\frac{1}{r_{mn,p}^3} \left(\delta_{ij} - 3\hat{r}_{mn,p}^{i}\hat{r}_{mn,p}^{j}\right)\,\text{,}
\end{equation}
where \YCY{$\vec{r}_{mn,p} = \vec{r}_{B,mn} - \vec{r}_{M,p}$}. 
Introducing the coefficient matrix simplifies the problem to matrix multiplication, allowing GPU acceleration to be utilized.

\YCY{
Three physical parameters obtained from the experimental setup are used in the calculations:\\
(1) B width and B height, which are the physical scales of the magnetic field map; \\
(2) M width and M height represent the physical scales of the magnetization source map;\\
(3) Shift vector $\vec{s} = \{s_x,s_y,s_z\}$ (green arrow) denotes the origin of the magnetization map relative to the magnetic field map plane. Here we assume the field plane and the magnetization source plane are parallel to each other, where $s_z$ corresponds to the sample-sensor distance (denoted as $d$). We allow $s_x$ and $s_y$ to be a small non-zero value since the two imaging focal planes in the quantum sensing microscope can be slightly shifted in x and y.
}

The magnetic inverse problem is ill-posed. Typically, one can easily reconstruct the magnetic fields, but the reconstructed magnetization can be drastically nonphysical \cite{lima2006magnetic}. To solve this problem, our model incorporates a parameterized physics-informed loss function:
\begin{equation} \label{loss}
    \begin{split}
        \text{Loss}&(\alpha,\beta,(\gamma,\text{Q}_0)) 
        \\
        &= \norm{\mathbf{A}\cdot\mathbf{m}-\mathbf{B}_{\text{exp}}}_{F}^{2} \\
        &+ \frac{\alpha}{2}\left(\norm{\Delta_x \mathbf{m}}_{2}^{2} 
        + \norm{\Delta_y \mathbf{m}}_{2}^{2}\right)\\
        &+ \beta \norm{\mathbf{m}}_{2}^{2} 
        + \gamma (Q - Q_0)^2
    \end{split}
\end{equation}
where $\norm{\mathbf{x}}_{F}^{2}$ is the square of Frobenius norm, and $\norm{\mathbf{x}}_{2}^{2}$ is the square $L^2$ norm. Each term is normalized to ensure a balanced contribution across all components. $\Delta_x{\mathbf{m}}$ and $\Delta_y{\mathbf{m}}$ represent the differences of the vector $\mathbf{m}$ along $x$-axis and $y$-axis on the discrete grid, respectively.

The first term of the loss function is the residual, which describes the difference between the experiment and the reconstructed magnetic fields. The $\alpha$ term favors continuous magnetization for physically meaningful solutions. In many condensed matter systems, the effective Hamiltonian tends to favor a continuous variation of spin directions \cite{newell1953theory,katsura1962statistical}. Even in the Heisenberg model with a negative exchange interaction J (representing anti-ferromagnetism systems), spin directions can change gradually when considering interactions beyond just nearest neighbors \cite{scholl2000observation,shpyrko2007direct}, leading to smoother transitions across the lattice. The $\beta$ term punishes large magnitudes of magnetization, resulting in a more average distribution, to avoid the physically unreasonable fluctuating distributions. Without $\beta$ regularization, the optimization might assign an nonphysically high value to a single pixel if this pixel can substantially reduce the error term, resulting in an incorrect representation. Typically, the $\alpha$ and $\beta$ terms act together, leading to faster convergence and smoother results.

Additionally, the tool allows for adding in prior assumptions of the real magnetization distributions in experiments. Here, we incorporate a topological $\gamma$ term used when solving magnetic spin textures \cite{yin2016topological,rybakov2019chiral}. A simple example is skyrmion, with topological charge $Q=1$, which can be calculated in the continuous form:
\begin{equation}
    Q = \frac{1}{4\pi}\int \mathbf{m}\cdot \left( \frac{\partial \mathbf{m}}{\partial x} \times \frac{\partial \mathbf{m}}{\partial y} \right) \dd x \dd y \, \text{,}
\end{equation}
or in the discrete form:
\begin{equation}
    Q = \frac{1}{4\pi} \sum_{x,y} \mathbf{m}_{\mathbf{r}}\cdot \left( \mathbf{m}_{\mathbf{r}+\hat{x}} \times \mathbf{m}_{\mathbf{r}+\hat{y}} \right)\, \text{.}
\end{equation}

\begin{figure*}[ht]
    \centering
    \includegraphics[width=1\textwidth]{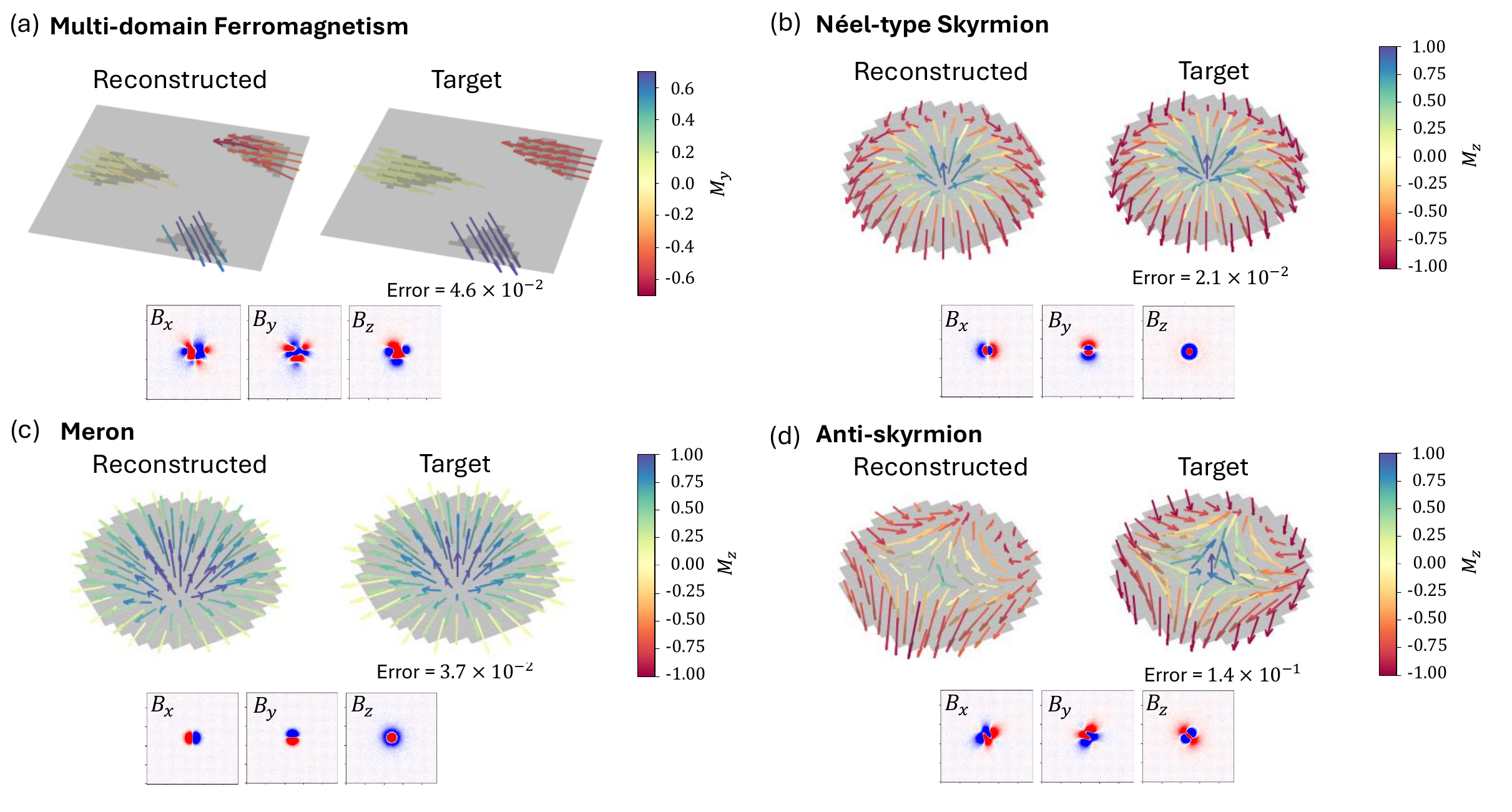}
    \flushleft
    \caption{Comparison between target and reconstructed magnetization for various spin configurations. The magnetic field was measured at $d=5$ with an optical resolution $w=2$ and an SNR of $5$ and shown below the magnetization. \YCYNEW{The arrow length is used to represent the magnitude of the dipole moment.} (a) Ferromagnetic regions, each with distinct randomly assigned uniform magnetization, separated by a distance of approximately 10. (b$\sim$d) Topological charge reconstruction for samples with a radius of 8. \YCYNEW{The color bar highlights the z component, providing a clearer representation of the topological charge structures.} (b) Néel-type skyrmion with topological charge $Q=1$. (c) Meron with $Q=1/2$. (d) Anti-skyrmion with $Q=-1$.}
    \label{fig:case}
\end{figure*}

In the reconstruction process, the error term in the loss function should dominate, and we need \Modified{physical regularization} parameters (\(\alpha\), \(\beta\), and \(\gamma\)) to guide the reconstruction without overshadowing the error term. In a reasonably noisy environment where the error term is typically around \(10^{-2}\), the regularization parameters are empirically selected to be significantly smaller to ensure optimal reconstruction performance. For instance, in Figure~\ref{fig:case}(a), \(\alpha\) and \(\beta\) are set to \(2\times10^{-5}\) for the multi-domain ferromagnetism case. Similarly, in Figure~\ref{fig:case}(b), \(\alpha\) and \(\beta\) are set to \(2\times10^{-5}\), while \(\gamma\) is \(8\times10^{-3}\) for the Néel-type skyrmion case with \(Q=1\). The reconstruction results remain robust against small variations in these empirical values.
\YY{In Appendix~B, we examine the impact of assuming an incorrect topological charge on the reconstruction results.}

Figure \ref{fig:diag}(b) shows the flow chart of our reconstruction model. It uses an experimentally measured vector magnetic field map, an assigned magnetization region, and defined spatial parameters as inputs and runs an optimization using the stochastic directional Newton's method~\cite{kandala2017eigen}. The process is then guided by the defined loss function, enabling the magnetization to rapidly converge to a physically meaningful result.

\section{Case Study}
In this section, we use a Néel-type skyrmion of radius $r=8$ as a demonstration. As shown in Figure \ref{fig:skyrmion}(a), the sample of radius $r$ is placed at a distance of $d$ from the sensor plane. In experiments, measurement noise and finite spatial resolution are inherent in the sensed field maps. 
\YY{To mimic these experimental conditions, we add Gaussian noise at a signal-to-noise ratio (SNR) of $5$ to model measurement noise.
We then simulate finite spatial resolution by convolving the map with a Gaussian kernel of full width at half maximum (FWHM) $w$. 
Parameters are normally expressed in physical units of pixel length from the experimental field map. In this case study, we instead present them as dimensionless values normalized to the pixel length.
}
The resulting field maps are then input into the model for reconstruction. As shown in Figure \ref{fig:skyrmion}(c), we define a reconstruction error 
\[
\text{Error} = \frac{\sum \left( \vec{\bm{m}}_{\text{target}} - \vec{\bm{m}}_{\text{recons.}} \right)^2}{\sum \vec{\bm{m}}_{\text{target}}^2}
\]
to quantify the fidelity of the reconstruction. 

The success of magnetization reconstruction is closely linked to the amount of magnetic information present in the field maps. Experimentally, the quality of this information is mainly influenced by three factors: noise level, spatial resolution, and the distance from the source. While noise levels in sensing experiments can typically be reduced through extended averaging, tuning distance and spatial resolution in an in-situ setting is often challenging. As shown in Figure \ref{fig:skyrmion}(d), we simulate the reconstruction process across various combinations of $w$ and $d$, revealing that for a given spatial resolution constraint, there is an optimal distance at which spin configuration imaging yields the best results. \YYNEW{Interestingly, when the magnetic field is measured at a close distance with a low-resolution sensor, the reconstruction quality deteriorates further. We attribute this degradation to localized features and steep spatial gradients in the near-source magnetic field, which become smeared out when undersampled, leading to significant loss of information.}

Our tool also supports physical \YCY{parameter optimization} for systems where \YCY{the shift vector $\vec{s}$ is} unknown or difficult to experimentally measure with enough precision. To verify the accuracy of the \YCY{parameter optimization}, we performed an additional simulation with the Néel-type case study.  Figure \ref{fig:skyrmion}(e) shows the results where all three \YCY{components of the shift vector} are fitted simultaneously. With each iteration, the parameters converge to the target value that we've assigned to represent unknown experimental parameters. Notably, the reconstruction error decreases with each iteration: reconstructed spin configuration becomes increasingly similar to the target configuration, demonstrating the successful information extraction from magnetic field measurements.

\YYNEW{Because the magnetic-field maps are large, the reconstruction is computationally intensive. We therefore leverage GPU acceleration to substantially reduce reconstruction time. In a representative skyrmion reconstruction, the stochastic Newton method with $2\times10^5$ iterations converges in around 19.5 minutes on a NVIDIA 4070 GPU, compared with 8.1 hours on a 16-core CPU --- a $\approx25\times$ speed-up. Our implementation is available as open-source code on GitHub~\cite{websitekey}.}

\section{Universality Demonstration}
In this section, we showcase the universal applicability of the developed model by applying it to diverse magnetization configurations, as shown in Figure \ref{fig:case}. Additionally, we demonstrate that by adjusting the topological parameters ($\gamma$), our model effectively captures a range of spin textures.
From Figure \ref{fig:skyrmion}(d), we gain a general understanding of the acceptable combinations of $w$ and $d$ needed for accurate magnetization reconstruction. Here, we set the experimental parameters to $w=2$, $d=5$, and SNR$=5$ for all simulations.  

\begin{figure*}[ht]
    \centering
    \includegraphics[width=0.99\textwidth]{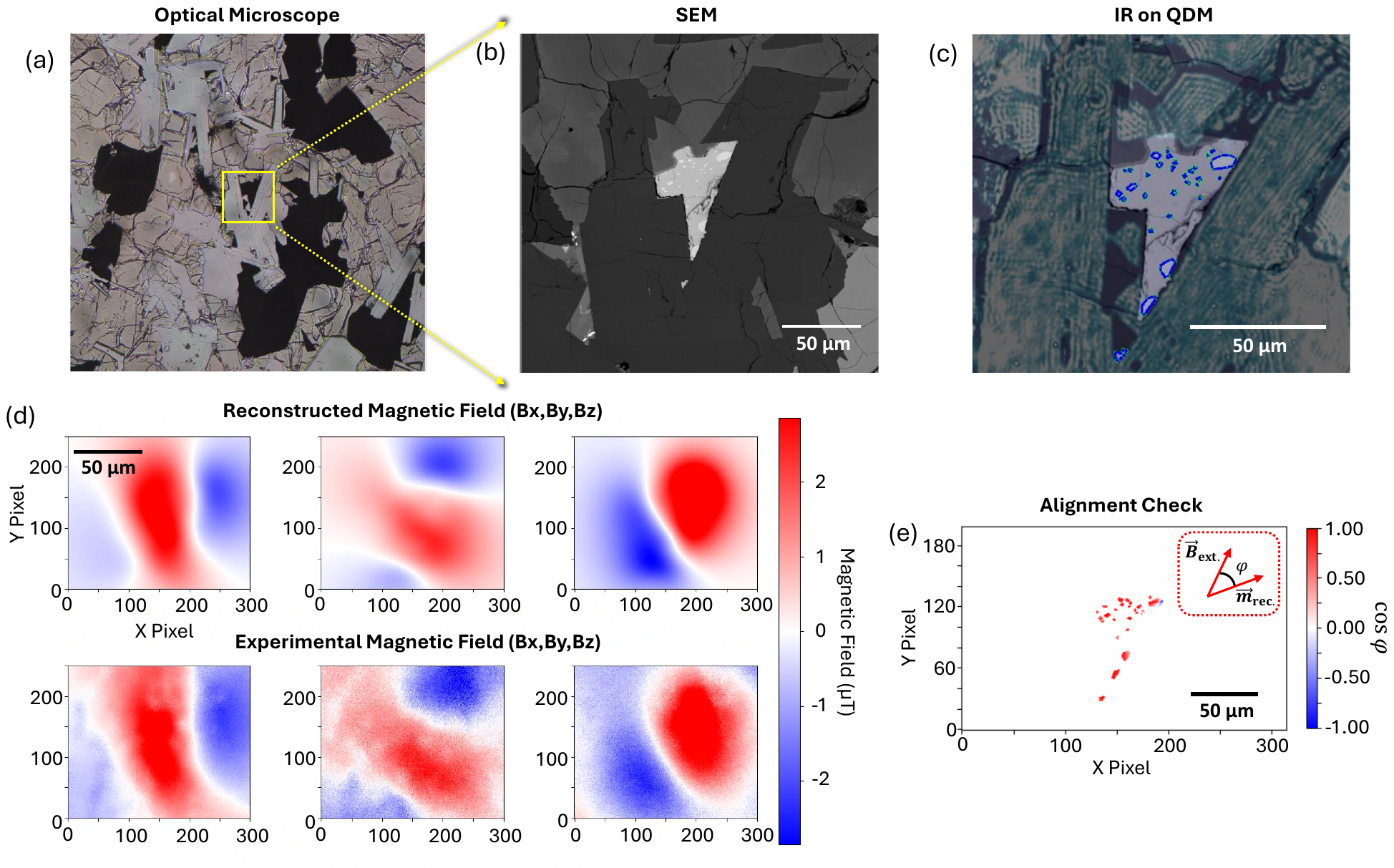}
    \flushleft
    \caption{(a) Optical microscope image of the moon rock sample 10003,184, with the yellow frame highlighting the region selected for subsequent SEM imaging. (b) Backscatterd SEM image showing the precise locations of potential magnetic rocks, visible as bright phases. (c) Infrared (IR) image acquired using the wide-field NV setup, covering the area where magnetic field measurements were performed. The IR image is overlaid on the SEM image, with the circled spots corresponding to the white spots in the SEM image, indicating the locations of magnetic sources (assigned magnetization region). (d) Comparison of experimental magnetic field data with reconstructed magnetic field components, displaying $B_x$, $B_y$, and $B_z$ from left to right. Measurements were conducted at room temperature with an external magnetic field applied in the direction of $(\text{-}1.07, \text{-}0.74, 0.65)$ mT. (e) $\cos{\varphi}$ map showing the alignment between the reconstructed magnetization vector and the expected direction (aligned with the applied field \YCYNEW{$B_\text{ext}$}). As shown in the inset, $\varphi$ is defined as the angle between reconstructed magnetization and the applied external magnetic field at each pixel.}
    \label{fig:rock}
\end{figure*}

Figure \ref{fig:case}(a) shows the reconstruction of a three-domain ferromagnetic sample, where each domain is assigned a random magnetization. This configuration is common in ferromagnetic materials \cite{taniuchi2016imaging}. The average distance between the domains is approximately 10. Despite some degree of cancellation from oppositely oriented magnetization domains, our model accurately reconstructs the magnetization. Figures \ref{fig:case}(b$\sim$d) depict the reconstruction of three distinct spin textures, each with a different topological charge. The spin texture region has a radius of 8. Figure \ref{fig:case}(b) shows a Néel-type skyrmion with a topological charge $Q=1$. Figure \ref{fig:case}(c) is a Meron with $Q=1/2$. Figure \ref{fig:case}(d) is an anti-skyrmion with $Q=-1$. These results highlight the versatility of our computational model, as it is able to adjust to different topological instructions by modifying the $Q$ values in the loss function. All reconstructions successfully capture the key features of the target spin configurations, demonstrating that our model is capable of reconstructing a wide range of source quantities. 

Notice that the anti-skyrmion reconstruction exhibits a higher error compared to the other three cases, primarily due to the violation of the $\beta$ condition in the anti-skyrmion configuration. While the anti-skyrmion maintains the same level of continuity as the skyrmion, its magnetic field contributions from different regions tend to cancel each other out to a larger extent. A magnetization distribution with smaller cancellations generates a magnetic field of relatively larger magnitude. To counteract this, the $\beta$ \YCY{regularization} guides the optimization toward a magnetization distribution with a smaller overall magnitude. This trade-off between the positive and negative effects of $\beta$ in this specific scenario ultimately results in an increase in the defined error.

\section{Experimental Demonstration}

\YYNEW{
The negatively charged NV center in diamond has a spin-1 ground state whose optical spin polarization and readout enables both wide-field and scanning magnetometry~\cite{casola2018probing,barry2020sensitivity,maletinsky2012robust,glenn2017micrometer,aslam2023quantum}. The local magnetic field projection onto the NV axis, $B_{\parallel}=\mathbf{B}\!\cdot\!\hat{\mathbf{n}}_{\rm NV}$, leads to a splitting between $|m_s = \pm1\rangle$, $2\gamma_e B_{\parallel}$, where $\gamma_e$ is the electronic spin gyromagnetic ratio.
This splitting can be experimentally obtained using optically detected magnetic resonance (ODMR) spectroscopy: by sweeping the frequency of the applied microwave drive while monitoring the fluorescence signal of NV centers, we expect a fluorescence drop when the microwave is resonant with an electronic spin transition $\ket{m_s{=}0}\!\leftrightarrow\!\ket{m_s{=}\pm1}$. Wide-field quantum diamond microscopy images a shallow layer of NV ensemble using a camera to yield a 2D map of $B_{\parallel}$ over large areas with optical diffraction-limit resolution, and four different crystal orientations of NV centers allow the reconstruction of a vector magnetic field map, $\mathbf{B}$. Scanning NV microscopy uses a single NV center on an atomic-force microscope (AFM) tip that brings the NV $\approx$~10-100 nm above the sample for nanoscale magnetic sensing. In this section, we validate our reconstruction accuracy using experimental data obtained from both wide-field and scanning NV magnetometry on known magnetization characteristics.
}

\begin{figure*}[ht]
    \centering
    \includegraphics[width=0.99\textwidth]{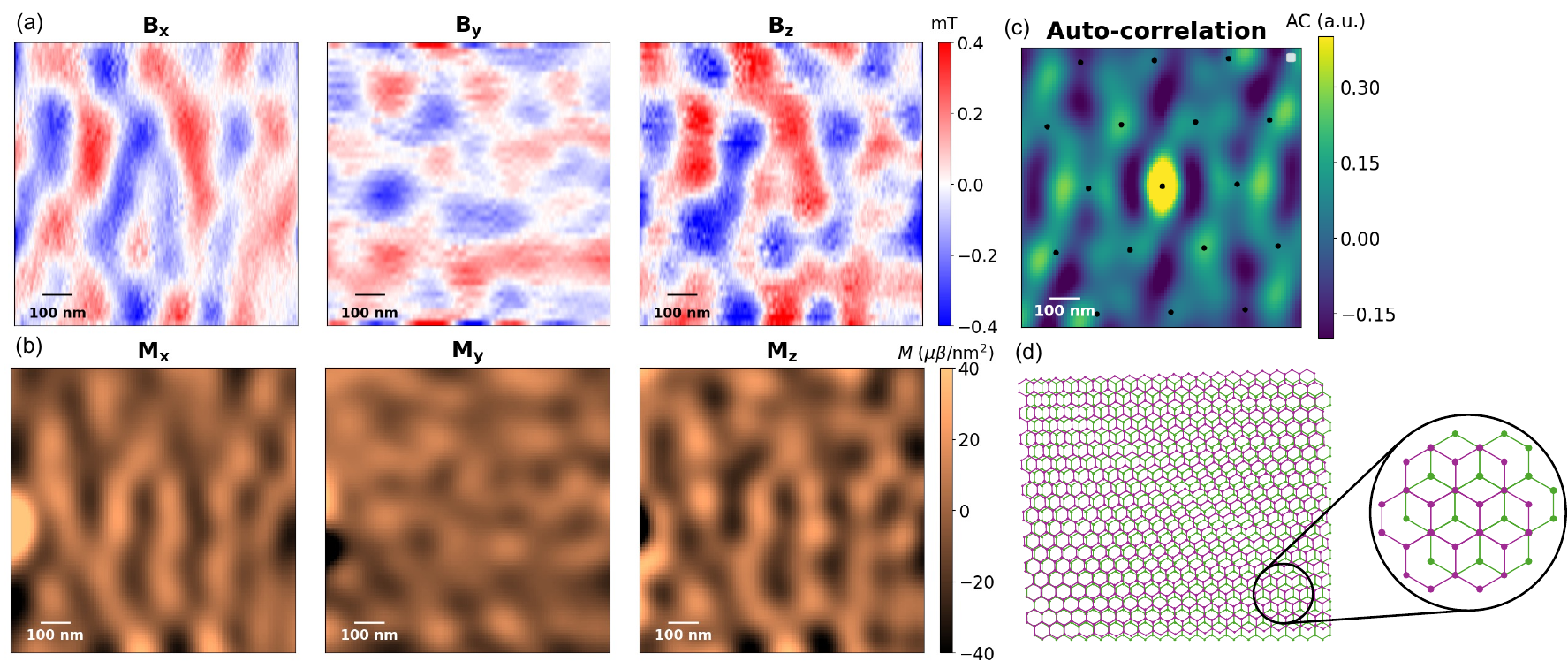}
    \flushleft
    \caption{\YCY{Magnetization reconstruction of 2D van der Waals material (\CrI). 
    (a) Magnetic field map along NV axis obtained using scanning NV magnetometer \cite{song2021moire}. Three-dimensional map is generated using method described in Appendix A~\cite{lima2009obtaining}.
    (b) Reconstructed surface magnetization. 
    (c) Three-dimensional auto-correlation map of the magnetization with lattice sites highlighted in black dots. 
    (d) Illustration of two \CrI\, layers twisted at a small angle generate a periodic stacking.}}
    \label{fig:moire}
\end{figure*}

We showcase the complete reconstruction process by applying our model to the magnetization reconstruction of the Apollo lunar mare basalt sample 10003,184. Lunar mineralogy and paleomagnetic research indicate that the predominant magnetic carriers in mare basalts are multi-domain metallic FeNi alloys, specifically kamacite (containing less than 5\% Ni) and martensite (5-25\% Ni) \Modified{which are considered carriers of stable remanent magnetization} \cite{WeissTikoo2014}. In this sample, we identify kamacite grains smaller than 50 µm, which are associated with troilite (FeS) in a eutectic assemblage, using a Scanning Electron Microscope (SEM). Transmission electron microscopy (TEM) analysis of these kamacite grains contains nanoscale sub-grains or revealed defects or nanoscale sub-grains \cite{Jung2024Assessing}. The sizes of these grains and sub-grains align with the single-domain (SD) to multi-domain (MD) spectrum \cite{Muxworthy2015}. Here, we identify these ferromagnetic species within the optical field of view, and we are able to generate magnetization results that align with the paleomagnetic understanding using vector magnetic field maps obtained through wide-field NV microscopy.

Figure \ref{fig:rock} presents both the optical (a) and backscatter SEM image (b) of magnetic minerals observed in our lunar sample. The bright phases in the backscatter SEM image are identified as kamacite grains, potentially containing defects or nanoscale sub-grains. These grains are also visible in the infrared-light optical image captured by the camera in our wide-field NV microscopy setup (Figure \ref{fig:rock}(c)). By targeting the same field of view, we are able to accurately map and pinpoint the precise locations of the magnetic sources (the circled areas). Continuous optically detected magnetic resonance (ODMR) measurements are then performed to obtain the local magnetic fields generated by the lunar rock sample near the diamond surface under an applied field \YCYNEW{$\vec{B}_\text{ext} = (\text{-}1.07, \text{-}0.74, 0.65)$mT}. 

In our experiments, lunar rock samples are glued into a 3-$mm$ thin section instead of being directly transferred to the diamond surface, making it difficult to measure the exact distance from the magnetic source to the NV plane. The focal planes for fluorescence and optical images often differ in NV sensing experiments, requiring z-axis adjustments to capture high-quality fluorescence images after locating the region of interest in the optical image, which will inevitably introduce some in-plane shifts. To address these challenges, we apply our parameter \YCY{optimization} and find x- and y-shifts of approximately $(12, 2)$ µm and an optimized distance of around $27$ µm, which aligns with our rough optical measurements.

Starting from the optimized experimental parameters, we conduct the \YYNEW{reconstruction} procedures. Figure \ref{fig:rock}(d) shows the near-perfect reconstruction of experimental magnetic field maps. Additionally, we demonstrate in Figure \ref{fig:rock}(e) that the reconstructed magnetization is well aligned with the applied field direction, consistent with our paleomagnetic expectation for these magnetic rocks.

\YY{
To further validate the robustness of our reconstruction framework on complex magnetic structures, we analyze the open-access data from Song \emph{et al.}~\cite{song2021moire}.
In that work, a scanning NV magnetometer is used to map the stray field above a twisted double-trilayer \CrI\ sample in which one three-layer stack is rotated by $0.3^{\circ}$ relative to the other.
This slight twist creates a \moire\ superlattice structure whose local stacking varies between monoclinic and rhombohedral registries---corresponding to interlayer antiferromagnetic (AFM) and ferromagnetic (FM) alignment, respectively. 
Because the intralayer AFM coupling within each trilayer largely cancels, domains with FM interlayer coupling dominate the measured signal. 
With a spatial resolution of $\lesssim50$~nm, the scanning NV magnetometer clearly resolves the magnetic fields from the moir\'{e} superlattice, which has a period of $150$~nm. }

\YY{
Figure~\ref{fig:moire}(a) shows the experimentally measured stray field map.
Since the scanning NV magnetometer can only measure the magnetic field projected along the NV axis, we recover the full vector field information using a previously developed Fourier transformation method~\cite{lima2009obtaining} (See Appendix~A.).
The reconstructed magnetization maps reveal rich nanoscale features in all three spatial components (Figure~\ref{fig:moire}(b)). 
Following the approach of Song \emph{et al.} \cite{song2021moire}, we compute the vector auto-correlation
\begin{equation}
\label{eq:autocorre}
    AC(\delta x, \delta y)
    = \langle
    \vec{M}(x+\delta x, y+\delta y) \cdot \vec{M}(x,y)
    \rangle_{x,y},
\end{equation}
}
\YCYNEW{
where \(\vec{M}_i(x,y)\) magnetization vector at grid $(x,y)$ to evaluate the reconstruction quality. }
\YY{
The resulting auto-correlation (Figure~\ref{fig:moire}(c)) reveals a clear hexagonal \moire\ pattern with a period $\sim 150~$nm, in excellent agreement with the structural superlattice constant (Figure~\ref{fig:moire}(d)).}

\begin{figure}[ht]
    \centering
    \includegraphics[width=0.49\textwidth]{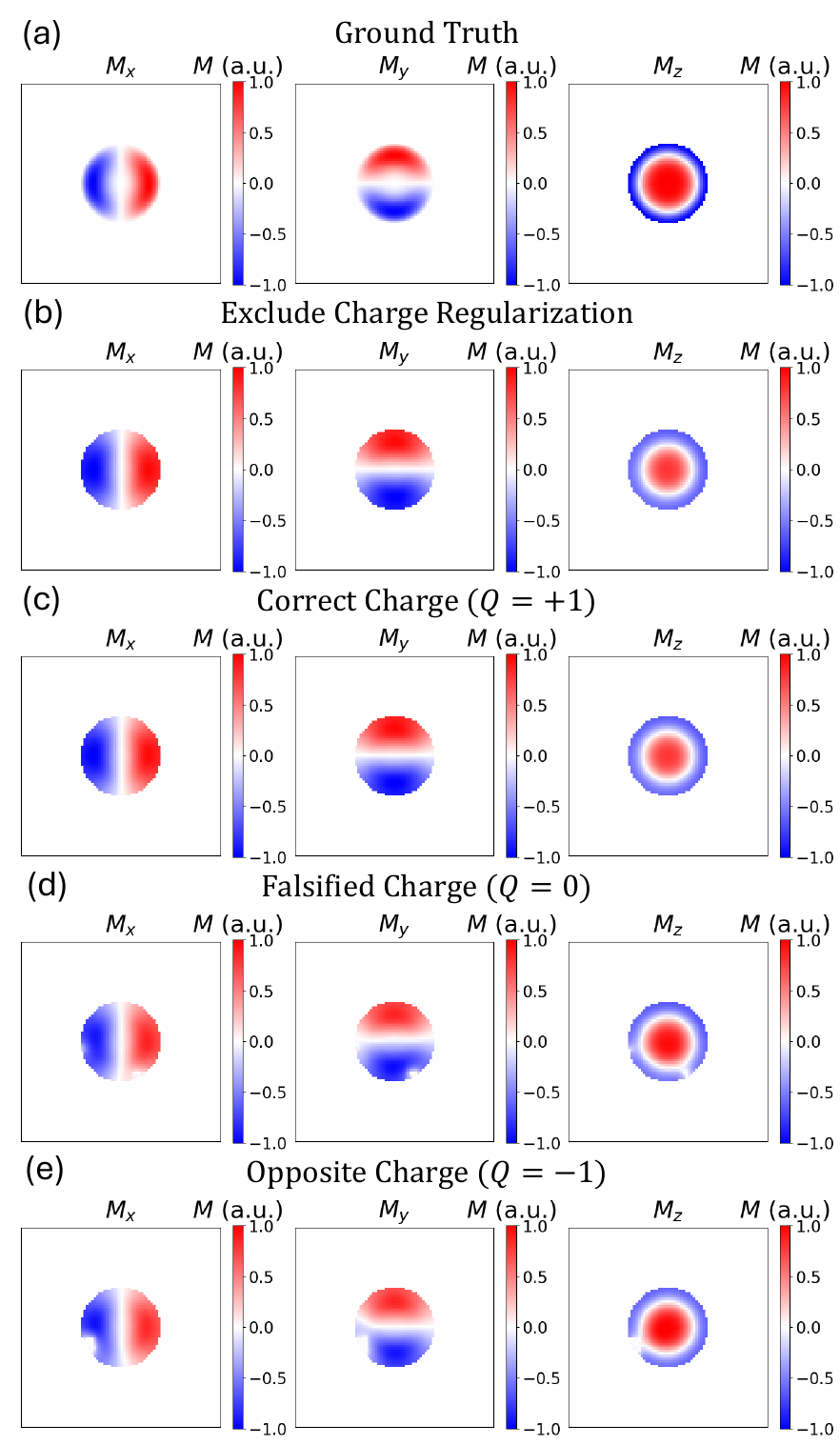}
    \flushleft
    \caption{\YY{Reconstruction of skyrmion magnetization under different assumptions of topological charge quantum number.}} 
    \label{fig:B}
\end{figure}

\section{Discussion}

In this work, we develop a computational model designed for the universal reconstruction of magnetization configurations from magnetic field maps. Our approach is structured to leverage GPU acceleration, incorporating a physics-informed, customizable loss function to ensure physically meaningful outcomes. Compared to existing methods, our approach offers clear advantages: it requires little to no prior knowledge of the magnetic source, can be customized for different levels of prior information, and enables rapid computation of the full vector magnetization across a two-dimensional plane.

This tool enables the accurate determination of spatial measurement parameters that are otherwise challenging to obtain experimentally. After optimizing these parameters, one can first run the algorithm without prior assumptions to achieve a reasonable initial result. Then, based on insights gained from this initial optimization, the loss function can be adjusted to refine the solution toward a physically meaningful outcome. Additionally, the tool allows for simulating magnetization reconstruction across different noise levels, spatial resolution limits, and sample-to-sensor distances, helping to identify the optimal experimental conditions. With these capabilities, the computational model offers significant values for a broad spectrum of magnetic imaging experiments across different fields of study.

\section{Acknowledgments}
\begin{acknowledgments}
We gratefully acknowledge Yun Yang and Bingtian Ye for helpful discussions. We thank Ruotian Gong, Guanghui He and Zijian Li for their assistance in the experiment. This work is supported by National Science Foundation under under Grant number 2514391 and 2152221 (NRT LinQ), and the Center for Quantum Leaps at Washington University. Y. Wang and C. Zhang are supported by National Science Foundation OSI-2503230, PHY-2409943, OSI-2228725, CNS-2211989.
\end{acknowledgments}

\section{Code Availability}
\YY{
The code for the developed toolset is available for download~\cite{websitekey}.}

\begin{figure*}[ht]
    \centering
    \includegraphics[width=0.7\textwidth]{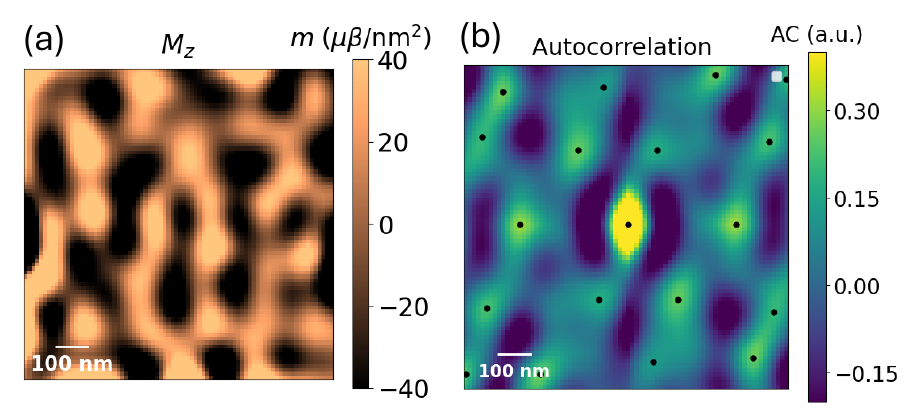}
    \flushleft
    \caption{\YY{Reconstructed lattice and auto-correlation analysis of \moire\ pattern under assumption of out-of-plane magnetization only \cite{song2021moire, park2025, huang2017layer}. 
    (a) Reconstructed magnetization under constraint that only $z$ component can contribute to the loss. $M_x$ and $M_y$ are suppressed to zero.
    (b) Auto-correlation map using Eq~\eqref{eq:autocorre}. Local maximum of auto-correlation are marked with black dots.
    }}
    \label{fig:C}
\end{figure*}

\section{Appendix A: Fourier Method}
\label{app:fourier}

\YCY{
The Maxwell equations impose strict constraints on any static, source-free magnetic field:
\[
    \nabla\!\cdot\mathbf B=0,
    \quad
    \nabla\!\times\mathbf B=0,
    \quad
    \mathbf B\to 0\quad (z\to\pm\infty).
\]
By restricting our measurement to the plane \(z=0\) and assuming no currents or magnetization above that plane, one can recover the full vector field \(\mathbf B(x,y,0)\) from its projection onto any axis \(\hat n\) that has a nonzero in-plane component.
}

\YCY{
In our reconstruction, we follow Lima \emph{et al.}\ \cite{lima2009obtaining}. Applying a 2D Fourier transform in \(x, y\) to Maxwell's equations gives
\begin{align}
  ik_y\,b_z(k_x,k_y,z) \;-\;\partial_z\,b_y(k_x,k_y,z) &= 0, 
  \label{eq:b1}\\
  ik_x\,b_z(k_x,k_y,z) \;-\;\partial_z\,b_x(k_x,k_y,z) &= 0, 
  \label{eq:b2}\\
  ik_y\,b_x(k_x,k_y,z) \;-\;ik_x\,b_y(k_x,k_y,z) &= 0,
  \label{eq:b3}
\end{align}
}
where \(b_i(k_x, k_y, z)\) is the Fourier component of \(B_i(x,y,z)\). 
Under the source-free, upward-decaying assumption, the vertical derivative is replaced by 
\begin{equation*}
  \partial_z \longrightarrow
  \begin{cases}
    -k & \text{if no sources lie above the } z=0 \text{ plane},\\
    +k & \text{if no sources lie below the } z=0 \text{ plane},
  \end{cases}
\end{equation*}
\begin{equation}
\label{eq:k_def}
 k \equiv \sqrt{k_x^2 + k_y^2} ,
\end{equation}
\YCY{
which implements the standard upward continuation \cite{blakely1996potential}.
}

\YCY{
If the measured vector field is the projection
\[
    B_n(x,y) = \mathbf B(x,y,0)\cdot\hat n,
\]
then its 2D FFT,
\[
    b_n(k_x, k_y) = n_x\,b_x+n_y\,b_y+n_z\,b_z,
\]
together with Eqs.~\eqref{eq:b1}-\eqref{eq:b3}, forms a linear system for \(b_x, b_y, b_z\).
An inverse FFT then yields the full planar vector map \((B_x, B_y, B_z)\), with only singular behavior at $k = 0$.
}

\section{Appendix B: Topological Charge Q}
\label{app:topo}

\YY{
The topological charge term can be incorporated into the total loss function to guide the reconstruction. 
In Fig.~\ref{fig:B}, reconstructing a single skyrmion domain with either zero weight ($\gamma=0$, panel b) or the correct charge ($Q=+1$, panel c) in the loss \eqref{loss} yields a magnetization map output that closely matches the ground truth (panel a). However, if $Q$ is incorrectly set to $0$ (panel d) or $-1$ (panel e), the resulting maps exhibit localized defects imposed by the topological constraint that cannot be eliminated by optimization.}

\YY{
As demonstrated, when the correct topological charge is known a priori---whether from theoretical considerations or independent measurement---the reconstruction reliably reproduces the true skyrmion magnetization. The model remains robust across physically valid topologies, while being sensitive to incorrect priors. Therefore, integrating precise topological constraints into the loss function is important for obtaining unbiased reconstruction results.
}

\section{Appendix C: Single Axis Constraint}
\YY{
During the optimization process, the stochastic directional Newton's method can be constrained to a single axis whenever the true magnetization alignment is known a priori. This is especially useful in cases where the moments are restricted to lie purely in-plane or purely out-of-plane: by enforcing an axial constraint, one can significantly speed up convergence.
}

\YY{
To explore this, we revisit our reconstruction of the twisted \CrI\ bilayer. Bulk and monolayer layer \CrI\ is known to prefer out-of-plane magnetization~\cite{song2021moire, park2025, huang2017layer}, so here we enforce $M_x=M_y=0$ and allow only $M_z$ to vary. The resulting magnetization map (Figure~\ref{fig:C}) has only a $z$-component, and its auto-correlation (Figure~\ref{fig:C}) recovers the same clear hexagonal \moire\ pattern seen in the full-vector case.
}

\bibliographystyle{apsrev4-2}
\bibliography{ref}
\end{document}